\documentclass[conference]{IEEEtran}
\IEEEoverridecommandlockouts
\usepackage{cite,subfigure}
\usepackage{amsmath,amssymb,amsfonts,amsthm}
\usepackage{algorithmic}
\usepackage{textcomp}
\usepackage{booktabs} % for specialrule
\usepackage{makecell} % for diaghead
\usepackage{xcolor}
\usepackage{pgf,pgfkeys,tikz,float,bm,caption}
\usepackage[ruled,linesnumbered,vlined]{algorithm2e}
\usepackage{graphicx,multirow}%
\usepackage{steinmetz}
\usepackage{siunitx,upgreek}
\def\BibTeX{{\rm B\kern-.05em{\sc i\kern-.025em b}\kern-.08em
    T\kern-.1667em\lower.7ex\hbox{E}\kern-.125emX}}
\graphicspath{{figures/}}

\begin{document}

\title{False Target Detection in OFDM-based Joint RADAR-Communication Systems}

\author{\IEEEauthorblockN{Antonios Argyriou%\textsuperscript{\orcidlink{0000-0002-2510-3124}}
	}
	\IEEEauthorblockA{Department of Electrical and Computer Engineering, University of Thessaly, Greece} 
}

\maketitle

\begin{abstract}	
	Joint RADAR communication (JRC) systems that use orthogonal frequency division multiplexing (OFDM) can be compromised by an adversary that re-produces the received OFDM signal creating thus false RADAR targets. This paper presents a set of algorithms that can be deployed at the JRC system and can detect the presence of false targets. The presence of a false target is detected depending on whether there is residual carrier frequency offset (CFO) beyond Doppler in the received signal, with a Generalized Likelihood Ratio Test (GLRT). To evaluate the performance of our approach we measure the detection probability versus the false alarm rate through simulation for different system configurations of an IEEE 802.11-based JRC system.
\end{abstract}


\section{Introduction}
\label{section:introduction}
Joint RADAR Communication (JRC), or Dual Function Radar Communication (DFRC) systems, are a subcategory of integrated sensing and communication (ISAC) systems where RADAR functionality and communication is implemented with the same waveform~\cite{Zhang21,Oliveira22}. JRC/DFRC systems are already a reality because they can leverage existing wireless digital communication protocols for building RADAR functionality on top of them~\cite{wisee13,Braun14,spotfi15,Kumari18,Li20}. 

When the JRC system uses orthogonal frequency division multiplexing (OFDM), it can leverage in the received signal, that has echoed from a target, data across the subcarriers to estimate the Range-Doppler response~\cite{Braun14,spotfi15,Kumari18}. Each specific subcarrier experiences a different phase offset for a certain delay of the signal, leading to a straightforward way of estimating range~\cite{Braun14} (e.g. with the DFT). On the other hand, Doppler can be estimated by using successive OFDM symbols and their returns (emulating thus a pulsed RADAR system)~\cite{book:fundamentals-of-radar-signal-processing}. 
Besides the ease of implementing ranging, another advantage is that OFDM wireless communication systems are used everywhere either in the omnipresent wireless LANs (802.11), or in mobile communications (5G). Thus, the signals transmitted from these OFDM-based systems offer an opportunity for embedding RADAR functionality into them at very low cost. Overall OFDM-based RADAR has been a relatively recent proposal~\cite{Braun14} with several applications being built around its capabilities~\cite{Li20,WiVi13,spotfi15,Lavery22}. 

\begin{figure}[!t]
	\centering
	\includegraphics[width=0.99\linewidth]{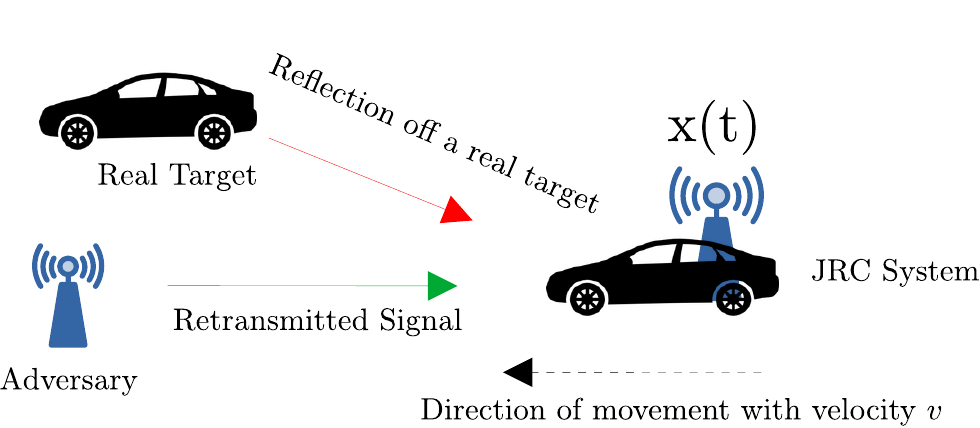}
	\caption{A possible scenario where the JRC system is deployed in a vehicle that transmits the OFDM signal $x(t)$. The adversary emits a signal (green) which is the retransmission of the received signal.} 
	\label{fig:system-ofdm-radar-detection}
\end{figure}

\textbf{Problem statement:} Despite its great appeal we argue that it is easy to deceive OFDM RADAR with a wireless device that uses the same communication physical layer. An adversary can use its wireless transmitter and emit not just typical jamming signals for preventing detection of real targets~\cite{Pirayesh22}, but a signal that leads the JRC RADAR algorithm to detect a false target. As an example, consider a scenario where the JRC system is deployed in a vehicle (Fig.~\ref{fig:system-ofdm-radar-detection}). Detecting an adversary that only acts as a jammer that transmits a power signal in such a scenario might be possible with machine learning techniques~\cite{jnl_2018_elsevier}. But detecting whether a received OFDM signal is from a real target or not is more challenging and in this case may lead to life-threatening decisions by the automatic driver assistance system (ADAS). In this paper we examine a wireless adversary that can use the same OFDM communication scheme with the JRC system, receive the signal, process it, and re-transmit it in the same frequency band in a way that mimicks a reflection from a target. Hence, the problem is to make OFDM RADAR more robust to false targets generated artificially by adversaries that are also wireless OFDM-capable devices.

\textbf{Proposed Approach:} In this paper we study this problem and we propose a system that leverages established OFDM RADAR algorithms and on top of them detects the false target. The basic intuition is that artificial signals generated by a wireless transmitter will contain impairments of the hardware,  and the most prevalent one is frequency variations beyond Doppler. In particular there will be a residual carrier frequency offset (CFO) between the baseband transmitted signal and the baseband received signal from a false target. This is because the local oscillator at the transmitter/receiver of the JRC OFDM-based RADAR will be different with the one at the adversary that generates the false target. In the case that the signal is a legitimate target, after downconversion at the OFDM RADAR receiver the echo will contain only the impact of Doppler affected by the target cross-section~\cite{book:fundamentals-of-radar-signal-processing}. Based on this intuition we first develop very detailed signal models to capture the discussed effects among others that are specific to OFDM signals. Then, we extend  the OFDM RADAR Range-Doppler derivation algorithm so that we can also estimate the unknown CFO. Our final step is to develop a Generalized Likelihood Ratio Test (GLRT) for the false target detection problem, and study it through simulations. Our results show that at minimal cost we can embed reliable false target detection functionality  in OFDM RADAR.

\section{Signal Models}

\subsection{Threat Model}
Fig.~\ref{fig:system-ofdm-radar-detection} depicts a representative scenario that illustrates the fundamental threat model under which the JRC system in this paper will operate. The adversary receives the JRC OFDM signal, demodulates it, remodulates it with OFDM, and re-transmits it so that at the JRC receiver it is perceived like an echo from a legitimate target. The JRC receiver processes the OFDM return signals with existing RADAR algorithms plus the proposed ones in this paper. The objective of the JRC system is to detect whether an adversary has re-transmitted the waveform, or the received OFDM signal is the reflection from a real illuminated  target (Fig.~\ref{fig:system-ofdm-radar-detection} top).

\subsection{Radar Path Loss Model}
We are interested in the RADAR functionality of the JRC system, and not its digital communication capability. Hence, we focus on the signal models that are used for RADAR processing. The large-scale radar channel gain for a target at distance $R_0$ is assumed to follow the free-space path-loss model set with a path loss exponent equal to 2~\cite{Bazzi12,Kumari18} and is given by:
\[
G=\frac{\lambda^2 \sigma_\text{RCS}}{64 \pi^3 R_0^4	}
\]
The radar cross section (RCS) of a real target is denoted as $\sigma_{\text{RCS}}$, and $\lambda$ is the wavelength.  The baseband complex channel gain for a single target is then:
\[
h= g\sqrt{G} 
\]
In the above  $g$ is a sample from a complex Gaussian random process that corresponds to Rayleigh fading that is constant (slow fading) for the duration of the RADAR coherent processing interval (CPI) (which is equal to the duration of a physical layer frame that consists of several symbols)~\cite{book:fundamentals-of-radar-signal-processing}. The passband signal model that we will present in the next section also takes into account the two-way delay of the signal, and Doppler.

\begin{figure}[t]
	\centering
	\includegraphics[width=0.65\linewidth]{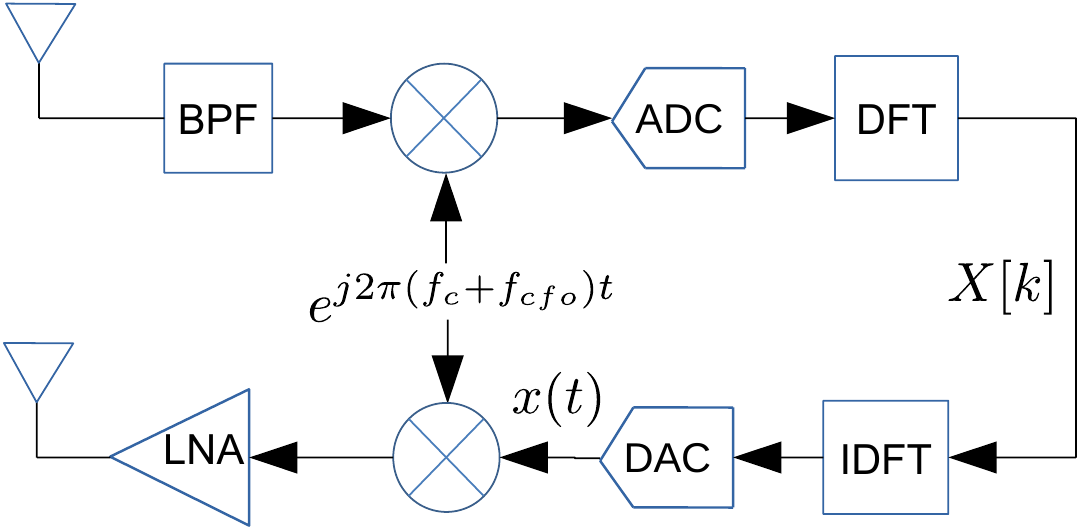}
	\caption{Adversary receiver (top) and transmitter (bottom) processing chain. }
	\label{fig:ofdm-radar-false-detection-adversary}
\end{figure}

\subsection{Signal Model for a False Target}
How can the JRC system model the adversary behavior? Different types of jammers and adversaries can be assumed requiring potentially different types of detection techniques~\cite{jnl_2018_elsevier}. In this paper the idea is to derive the OFDM signal model at the JRC receiver when the adversary creates a re-transmission of it. The adversary does not reflect the signal but it demodulates it and remodulates adding thus an additional constant system delay (Fig.~\ref{fig:ofdm-radar-false-detection-adversary}). To make the notation consistent we denote this two-way plus system delay equal to $\frac{2R_0}{c}$ (i.e. $R_0$ is the equivalent distance that would generate a delay $\frac{2R_0}{c}$ and is higher than the actual physical distance). The JRC system is not interested to make this distinction, leading thus to a single parameter $R_0$ capturing both. In Fig.~\ref{fig:system-ofdm-radar-detection} we observe that this two-way delay of the signal that includes also movement at velocity $v$ is time-varying and equal to $\tau(t)$=$\frac{2R_0}{c}-\frac{2vt}{c}$.

Now let $x(t)$ be the OFDM signal that the JRC transmitter emitted. We assume that the result of demodulation and remodulation by the adversary as depicted in Fig.~\ref{fig:ofdm-radar-false-detection-adversary} is perfect. This is something not unrealistic in WiFi OFDM communication in short distances. Even if this is not the case the interested reader is referred to ~\cite{cnf_2023_radarconf1} regarding the impact of imperfect demodulation of $x(t)$ on the OFDM RADAR Range-Doppler response algorithms. As seen in Fig.~\ref{fig:ofdm-radar-false-detection-adversary} this baseband signal $x(t)$ at the adversary is upconverted to the carrier frequency $f_c$ by going through the QAM modulator (mixing with $e^{j2\pi f_c t}$) but there is a residual CFO $f_\text{cfo}$ with respect to the JRC receiver which is locked at $f_c$. Hence, at the input of the JRC receiver the passband continuous time signal model in a narrowband flat fading (single-tap model with complex gain $h$) channel becomes:
\begin{align}
y_\text{PB}(t)=hx(t-\tau(t))e^{j2\pi(f_\text{cfo}+f_c)(t-\tau(t))}+w(t)\label{eqn:signal-model-wo-ofdm-narrowband}
\end{align}
In the above $w(t)$ is the AWGN sample. In our model we ignore the sampling clock offset (SCO) since we are not demodulating data symbols at the JRC system (it knows the symbols that it transmitted).

As we said the baseband signal $x(t)$ is the result of multi-carrier modulation and more specifically OFDM. With $N$ sub-carriers spaced at locations $f_k$=$k \Delta f$ \SI{}{\hertz} that can contain data, pilot symbols, or a combination of both (depending on the standard), the desired OFDM symbol in continuous time is:
\begin{equation}
	x(t)=\frac{1}{\sqrt{N}}\sum_{k=0}^{N-1}X[k]e^{j2\pi k \Delta f t},~~0\leq t \leq T_N
	\label{eqn:ct-ofdm} 
\end{equation}
$X[k]$ is the complex symbol modulated onto subcarrier $k$, and $T_N=N/\Delta f$ is the OFDM symbol duration. 

We also write the sampled version of the OFDM signal that we will need when we describe OFDM RADAR processing. By sampling the last equation with a rate of $f_s$ samples/sec at times $t=n/f_s$, and recalling that the fraction of the subcarrier frequency relative to the sampling rate, i.e. $f_k/f_s=k/N$, we get a digital frequency $k/N$ and the discrete form:
\begin{align}
	x[n] &=\frac{1}{\sqrt{N}}\sum_{k=0}^{N-1}X[k]e^{j2\pi n k/N}=\frac{1}{\sqrt{N}}\sum_{k=0}^{N-1}X[k]e^{j2\pi n k\frac{\Delta f}{f_s}}, \nonumber\\
	&~~0\leq n \leq N-1	\label{eqn:idft-ofdm} 
\end{align}
This is the inverse DFT (IDFT) of $X[k]$ allowing thus its well known efficient implementation also at the receiver.

To continue the derivation we have to combine~\eqref{eqn:signal-model-wo-ofdm-narrowband} and \eqref{eqn:ct-ofdm}:
\begin{align}
y_\text{PB}(t)=\frac{h}{\sqrt{N}}\sum_{k=0}^{N-1}X[k]e^{j2\pi (k\Delta f+f_c+f_\text{cfo})(t-\tau(t))} %\nonumber\\
+w(t)\label{eqn:signal-model-with-ofdm}
\end{align}
To get our final\textit{ discrete baseband model} we proceed as follows. The JRC receiver downconverts to baseband the signal in~\eqref{eqn:signal-model-with-ofdm} by multiplying with $e^{-j2\pi f_c t}$. In the resulting expression first we substitute $\tau(t)$=$\frac{2R_0}{c}-\frac{2vt}{c}$. Next, and since the JRC RADAR is interested in the range and Doppler we must write the discrete samples of the signal indexed by slow and fast time in 2D form~\cite{book:fundamentals-of-radar-signal-processing}. So we sample this analog signal at $t=n/f_s+mT_N$, where $m$ indicates the OFDM symbol (slow time sample). After doing the previous two steps a rather long expression of products of exponential terms emerges which is simplified as we describe next. First, we assume that the CFO and Doppler within an OFDM symbol causes negligible phase offset and so $e^{j2\pi (f_c+f_\text{cfo}+k\Delta f)\frac{2v}{c}n/f_s} \rightarrow 1$~\cite{Braun14}. Second, we notice that $e^{j2\pi k \Delta f mT_N}$=1. Third, there is a term $e^{-j2\pi (f_c+f_\text{cfo})\frac{2R_0}{c}}$ that is constant phase (i.e. it is not indexed by either $n,k$ or $m$). This means that it can be merged with $h$ to be jointly denoted as $h^{'}$. After these steps we have that the 2D sampled baseband received signal is:\footnote{We maintain the square root of $N$ since it will be remove after the IDFT.}
\begin{align}
	y[m,n] &=\frac{h^{'}}{\sqrt{N}} \sum_{k=0}^{N-1}X[k]e^{j2\pi  k\Delta f(\frac{n}{f_s}-\frac{2R_0}{c}+\frac{2v}{c}mT_N)} \nonumber\\
	&\times
	e^{j2\pi \big ((f_c+f_\text{cfo})\frac{2v}{c}mT_N+f_\text{cfo}(\frac{n}{f_s}+mT_N) \big ) } 
		\nonumber\\
	&+w[n],~~0\leq k,n \leq N-1,~~0\leq m \leq M-1
	\label{eqn:2d-sampled-baseband-received1}
\end{align}
This expression gives the final signal model for each subcarrier $k$ in terms of fast time indexes $n$ and slow time indexes $m$ (for each OFDM symbol $m$ of duration $T_N$). The total number of OFDM symbols contained in this 2D signal is $M$. 

An important difference with respect to other works on OFDM RADAR must be stated. In the above the Doppler effect is captured in the aggregate term $e^{j2\pi (k\Delta_f+f_c+f_\text{cfo})\frac{2v}{c}mT_N}$ for the $k$-th subcarrier, which means that we assume different Doppler for each subcarrier. In existing OFDM RADAR systems this is not the typical assumption, i.e. the assumption is the same Doppler for each subcarrier which means we would remove the term $k\Delta f$ from the last expression.

\subsection{Signal Model with a Real Target}
We must now derive a second signal model at the JRC receiver for the case that a real target reflects the signal. We can derive this signal model from the one before since now it is less complex. The differences are: The two way delay now that depends on the actual distance $R_0$, and the CFO is non-existent. Hence:
\begin{align}
&	y[m,n]=\frac{h^{'}}{\sqrt{N}} \sum_{k=0}^{N-1}X[k]e^{j2\pi  k\Delta f(\frac{n}{f_s}-\frac{2R_0}{c}+\frac{2v}{c}mT_N)} \label{eqn:2d-sampled-baseband-received2}\\
	&\times e^{j2\pi f_c\frac{2v}{c}mT_N  }+w[n],~0\leq k \leq N-1,~0\leq m \leq M-1\nonumber	
\end{align}
The models in~\eqref{eqn:2d-sampled-baseband-received1} and~\eqref{eqn:2d-sampled-baseband-received2} pave the way for our detection strategy.

\section{Range Doppler Estimation in OFDM RADAR}
We now briefly describe the basic algorithm behind OFDM RADAR for range and velocity estimation. The discussion is based on simpler signal models than \eqref{eqn:2d-sampled-baseband-received1} and~\eqref{eqn:2d-sampled-baseband-received2}.

\subsection{OFDM RADAR Assumptions}
First it is prudent to discuss assumptions that are specific to OFDM RADAR. In the general case it might be difficult for a JRC system to estimate velocity/range from an OFDM signal that is not designed specifically for RADAR. To meet certain RADAR requirements like range resolution, maximum unambiguous range, and velocity resolution, the waveform must have certain values for its parameters. For example range resolution is proportional to the signal bandwidth, i.e. $\Delta R=\frac{c}{2}BW$~\cite{book:fundamentals-of-radar-signal-processing}. Hence, for an OFDM signal with a total bandwidth of $N\Delta f$ this is equal to $\Delta R=\frac{c}{2N\Delta f}$ and is thus fixed for a specific digital communication standard. Similarly for OFDM velocity resolution is $\Delta v=\frac{c}{2MT_Nf_c}$~\cite{Braun14}. However, under certain but not limiting assumptions it is possible. 

These assumptions are: 1) The cyclic prefix (CP) duration is larger than the round-trip delay of the transmitted RADAR waveform. 2) The subcarrier distance is at least one order of magnitude larger than the largest occurring Doppler shift. 3) The transmitter distance remains constant during the transmission of one group of OFDM symbols that will be jointly processed (the well-known stop-n-hop approximation~\cite{book:fundamentals-of-radar-signal-processing}).

The second assumption is the most tricky to satisfy and depends on the standard.  However, if we consider 802.11a as an example the default subcarrier spacing is \SI{312.5}{\kilo\hertz} and with a \SI{20}{\mega\hertz} channel, we have that a $f_c$=\SI{5}{\giga\hertz} results in a Doppler shift of $f_D$=$16.6 v$~\SI{}{\hertz} which way more than an order of magnitude smaller than \SI{312.5}{\kilo\hertz} for any realistic velocity $v$. For higher $f_c$ (e.g. in the \SI{24}{\giga\hertz} ISM band) better resolution could be achieved but the previous upper bound is reduced.

\subsection{OFDM RADAR Algorithm}
\label{section:ofdm-radar-algorithm}
As in every receiver, the JRC RADAR receiver synchronizes to the start of the first wireless communication frame until the expected number of $M$ OFDM symbols is received for joint processing. With the standard OFDM RADAR algorithm the same Doppler across the subcarriers is assumed. This means that we can set $k\Delta f\frac{2v}{c}mT_N \rightarrow 0$ in~\eqref{eqn:2d-sampled-baseband-received2} to obtain a simpler expression for the return signal of an actual target:
\begin{align}
		y[m,n]&=\frac{h^{'}}{\sqrt{N}} \sum_{k=0}^{N-1}X[k]e^{j2\pi  k\Delta f(\frac{n}{f_s}-\frac{2R_0}{c})}e^{j2\pi f_c\frac{2v}{c}mT_N  } \nonumber\\
	&+w[n],~0\leq k \leq N-1,~0\leq m \leq M-1\label{eqn:2d-sampled-baseband-received3}
\end{align}
When the signal is received at the OFDM RADAR receiver, after downconversion and sampling and the signal is \eqref{eqn:2d-sampled-baseband-received3}, the next step is to process the signal with a DFT. To understand how we will get the next result note from~\eqref{eqn:idft-ofdm} that:
\begin{align}
\text{DFT} \Big ( \frac{1}{\sqrt{N}}\sum_{k=0}^{N-1}X[k]e^{j2\pi k\Delta f( \frac{n}{f_s}-\frac{2R_0}{c})} \Big ) =X[k] e^{-j2\pi k\Delta f\frac{2R_0}{c}} \nonumber
\end{align}
Besides the DFT of the $n$-indexed samples of the simplified 2D sample matrix in~\eqref{eqn:2d-sampled-baseband-received3}, we have to divide the result by $X[k]$ (to remove the known digital communication signal) to obtain:  
	\begin{align}
			\tilde{Y}[k,m]&=h^{'} e^{j2\pi f_c\frac{2v}{c}mT_N} e^{-j2\pi k\Delta f\frac{2R_0}{c}} +W[n], \nonumber \\
		& 0\leq k \leq N-1,~~0\leq m \leq M-1
	\end{align}
For this data model we then execute a 2D-DFT across the $m,k$ indexes with period $T_N$ and $\Delta f$ respectively. The resulting Range-Doppler response will exhibit peaks at locations $f_c\frac{2v}{c}$ and $\frac{2R_0}{c}$ allowing thus velocity and range estimation.

\section{CFO, Range, and Velocity Estimation and False Target Detection}
To detect the presence of a false target we adopt the Generalized Likelihood Ratio Test (GLRT). The GLRT is used when we have to test for two hypotheses but the two data models have unknown parameters. Hence, we must first estimate the unknown parameters under two signal models and then based on these estimates develop a threshold test so as to detect which of the two signal models is more likely to have occurred~\cite{kay98}. We first discuss OFDM RADAR processing of the downconverted and 2D sampled signal $y[m,n]$ at the JRC receiver chain (illustrated in Fig.~\ref{fig:ofdm-radar-false-detection-receiver}), which is slightly more complicated than the one we discussed in the last section. Then, we address the estimation of the unknown parameters $R_0,v$ under the hypothesis that the target is real, and $R_0,v,f_\text{cfo}$ under the hypothesis that the target is false. Finally, we develop the GLRT threshold test for detecting real targets.

\subsection{Signal Model after OFDM RADAR Preprocessing}
Unlike Section~\eqref{section:ofdm-radar-algorithm} where we noticed that it is easy to estimate range and velocity, with the signal models developed in this paper it is not. Let us see what happens under the two cases, namely false target and real target.

\subsubsection{False Target Present}
For the case of a false target if we perform the first necessary DFT in expression \eqref{eqn:2d-sampled-baseband-received1} at the OFDM RADAR receiver we have:
\begin{align}
	\text{DFT}(y[m,n])&=Y[k,m]=h^{'} X[k] e^{j2\pi ((f_c+f_\text{cfo})\frac{2v}{c}+f_\text{cfo})mT_N}\nonumber\\
	&\times e^{-j2\pi k\Delta f\frac{2R_0}{c}} e^{j2\pi k\Delta f \frac{2v}{c}mT_N}+W[n],\nonumber\\
	&0\leq k \leq N-1,~~0\leq m \leq M-1
	\label{eqn:signal-model-discrete-pre-process}
\end{align}
Then, the receiver at the JRC system divides~\eqref{eqn:signal-model-discrete-pre-process} by $X[k]$ to remove the known digital communication signal.\footnote{Note that in practice since the adversary has demodulated and re-modulated the symbols there might be errors especially in the low SNR regime.} Based on this the previous becomes:
\begin{align}
	\tilde{Y}[k,m]&=h^{'} e^{j2\pi ((f_c+f_\text{cfo})\frac{2v}{c}+f_\text{cfo})mT_N}  \nonumber\\
&\times e^{-j2\pi k\Delta f\frac{2R_0}{c}} e^{j2\pi k \Delta f\frac{2v}{c}mT_N}+W[n], \nonumber \\
& 0\leq k \leq N-1,~~0\leq m \leq M-1
	\label{eqn:signal-model-discrete-pre-process2}
\end{align}
Similar to a typical OFDM-based RADAR, at the JRC/DFRC system we perform a 2D-DFT on the signal across the $k,m$ indexes to obtain the Range-Doppler response. 

\begin{figure}[t]
	\centering
	\includegraphics[width=0.99\linewidth]{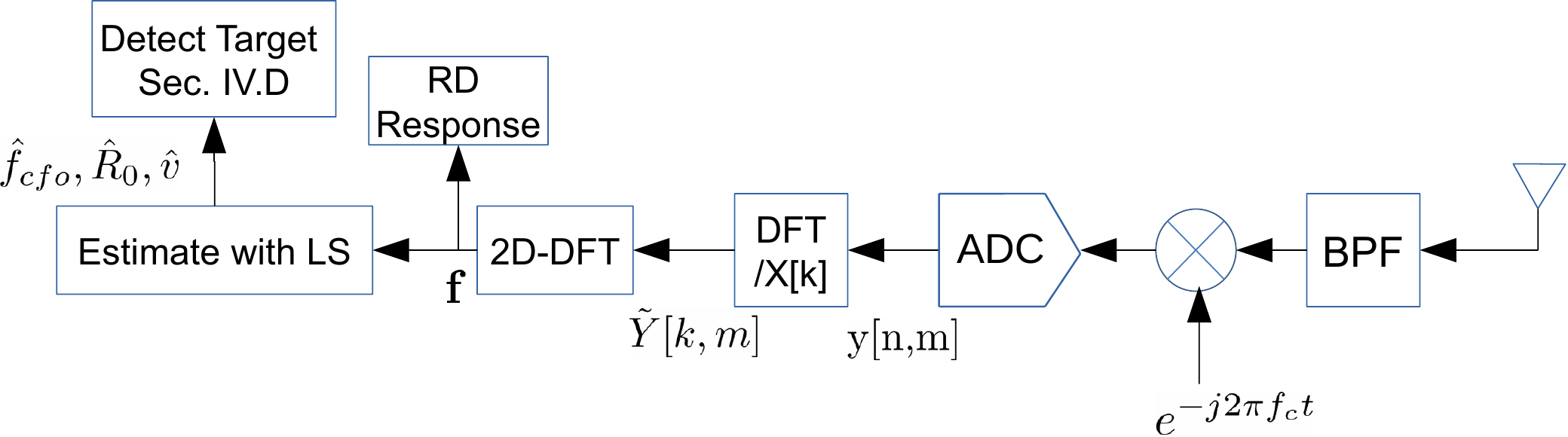}
	\caption{OFDM RADAR receiver processing chain highlighting the stage at which the proposed estimation \& detection algorithms are used.}
	\label{fig:ofdm-radar-false-detection-receiver}
\end{figure}

\subsubsection{Real Target Present}
Similarly to the last paragraph we have for the case that a real target exists:
\begin{align}
		\tilde{Y}[k,m]&=h^{'} e^{j2\pi f_c\frac{2v}{c}mT_N} e^{-j2\pi k\Delta f\frac{2R_0}{c}} e^{j2\pi k \Delta f\frac{2v}{c}mT_N} \nonumber\\
	& +W[n],~0\leq k \leq N-1,~0\leq m \leq M-1
	\label{eqn:signal-model-discrete-pre-process4}
\end{align}

\subsection{Range-Doppler Response}
The OFDM RADAR processing described in Section~\ref{section:ofdm-radar-algorithm} that employs a 2D-DFT cannot be used with the new developed data models in~\eqref{eqn:signal-model-discrete-pre-process2} and~\eqref{eqn:signal-model-discrete-pre-process4}: For a given $m$ that we have a constant phase, that may either include CFO or not, we calculate the DFT of~\eqref{eqn:signal-model-discrete-pre-process2} across the index $k$ and with sampling period $\Delta f$. It is easy to see that there will be $M$ peaks at $-\frac{2R_0}{c}+\frac{2v}{c}mT_N,~m=1,...,M$. Regarding the phase it can be estimated similarly by taking the DFT of~\eqref{eqn:signal-model-discrete-pre-process2} across the $m$ index (for a fixed $k$) and with DFT sampling period $T_N$ leading us to $N$ peaks at $\frac{2v}{c}(f_\text{cfo}+f_c)+f_\text{cfo}+\frac{2v}{c}k\Delta f,~k=1,...,N$. The same holds for~\eqref{eqn:signal-model-discrete-pre-process4} but without the CFO. As a result after taking the 2D-DFT on our data models in~\eqref{eqn:signal-model-discrete-pre-process2} and~\eqref{eqn:signal-model-discrete-pre-process4} there are multiple peaks in the Range-Doppler response. This is a result of the assumption that we do not have the same Doppler across the subcarriers. Hence, a second step is needed before estimating range and Doppler.

\subsection{Estimation of the Complex Sinusoidal Parameters from the Range-Doppler Response}

\textbf{False Target:} The previous peaks can be collected and we have the following linear data models (that include frequency domain AWGN denoted by $W[k],W[m]$):

\small
	\begin{align}
		&f_{peak}[m]=-\frac{2R_0}{c}+\frac{2v}{c}mT_N+W[m],0\leq m \leq M-1		\label{eqn:signal-model-discrete-pre-process3}\\
		&f_{peak}[k]=\frac{2v}{c}(f_\text{cfo}+f_c)+f_\text{cfo}+\frac{2v}{c}k\Delta f +W[n],0\leq k \leq N-1\nonumber
	\end{align}
\normalsize
As a result at this point we have a signal model with $M+N$ observations. We merge the previous two vectors, and we have the following signal model for estimation:
	\begin{align}\label{eqn:signal_model}
		\mathbf{f}=\mathbf{a}_1f_\text{cfo}+\mathbf{A}_2[R_0\quad \quad v]^T+\mathbf{W}
	\end{align}
Where the $(M+N)\times 2$ matrix $\mathbf{A}_2$, and $(M+N)\times 1$ vector $\mathbf{a}_1$ are trivially populated from~\eqref{eqn:signal-model-discrete-pre-process3}. With least squares (LS) we estimate $\hat{f}_\text{cfo}, \hat{R}_0,\hat{v}$.

\textbf{Real Target:} Under the real target present hypothesis we have the same model as in ~\eqref{eqn:signal_model} after we set $a_1 \rightarrow 0$ since there is no CFO under this case:
\begin{align}\label{eqn:signal_model_ls}
	\mathbf{f}=\mathbf{A}_2[R_0\quad \quad v]^T+\mathbf{W}
\end{align}
To summarize, after this stage we have estimates under the two hypotheses and we can proceed to detection, that is the next stage in the receiver processing chain (Fig.~\ref{fig:ofdm-radar-false-detection-receiver}).

\subsection{Detection for Constant CFO}
Recall that we have assumed that the CFO does remain constant for the complete $M\times N$ samples. Now our task is to detect the presence of a real target which means that we have to condition the pre-processed signal $\tilde{Y}[k,m]$ on whether a real target is present in the signal (hypothesis $\mathcal{H}_1$) or a false target is present (hypothesis $\mathcal{H}_0$). To make the expressions more compact we first vectorize $\tilde{Y}[k,m]$ into a $M\times N$ vector:
\[
\mathbf{z}[k+mK]=\tilde{Y}[k,m],\quad k=1,...,N, m=0,...,M-1.
\]
In the last subsections we have estimated three parameters under two hypotheses. With these estimates we re-create for a given OFDM symbol the signal of interest under the two hypotheses, by using ~\eqref{eqn:signal-model-discrete-pre-process2} and~\eqref{eqn:signal-model-discrete-pre-process4}:
%\small
\begin{align*}
\hat{\mathbf{u}}_{1}&= e^{j2\pi f_c\frac{2\hat{v}}{c}mT_N} e^{-j2\pi k\Delta f\frac{2\hat{R}_0}{c}} e^{-j2\pi k \Delta f\frac{2\hat{v}}{c}mT_N},\\
\hat{\mathbf{u}}_{0}&=e^{j2\pi ((f_c+\hat{f}_\text{cfo})\frac{2\hat{v}}{c}+\hat{f}_\text{cfo})mT_N} e^{-j2\pi k\Delta f\frac{2\hat{R}_0}{c}} e^{-j2\pi k \Delta f\frac{2\hat{v}}{c}mT_N},\\
k&=1,...,N, m=0,...,M-1.
\end{align*}
\normalsize
Note that these hypotheses remind us of the well known FSK detection problem in digital communication where we have to detect one of two frequencies. Only in this case we have not pre-selected the frequencies as in FSK. The two hypotheses are compactly modeled as:
\begin{align}\label{eqn:est}	
	&\text{under}~\mathcal{H}_0: \mathbf{z}=h^{'} \hat{\mathbf{u}}_0+\mathbf{w}\\
	&\text{under}~\mathcal{H}_1:\mathbf{z}=h^{'}\hat{\mathbf{u}}_1+\mathbf{w}
\end{align}
Regarding the noise covariance matrix of the vector $\mathbf{w}$ recall that it is positive semi-definite, i.e. $\bm{\Sigma}\succeq 0$. The PDF of the complex data $\mathbf{z}$ under $\mathcal{H}_0$ is
\begin{align}\label{eqn:fy_A}
	f(\bm{z}|\mathcal{H}_0)&=\frac{\exp [ -\frac{1}{2} (\mathbf{z}-h^{'}\mathbf{u}_0)^H\bm{\Sigma}^{-1}(\mathbf{z}-h^{'}\mathbf{u}_0) ] }{\sqrt{(2\pi)^{M+K} \det(\bm{\Sigma})}},
\end{align}
and similarly for $\mathcal{H}_1$. When we have uncorrelated noise with constant variance $\sigma^2$ the PDF of the data for the two hypotheses is:

\small
\begin{align}
	& f(\bm{z}|\mathcal{H}_0)=\frac{\exp \Big ( \frac{-\|\mathbf{z}-h^{'}\mathbf{u}_0\|_2^2 }{2\sigma^2} \Big ) }{(2\pi(\sigma^2))^{\frac{M+K}{2} }},f(\bm{z}|\mathcal{H}_1)=\frac{\exp \Big ( \frac{-\|\mathbf{z}-h^{'}\mathbf{u}_1\|_2^2 }{2\sigma^2} \Big ) }{(2\pi(\sigma^2))^{\frac{M+K}{2} }}\nonumber
\end{align}
\normalsize
After the signal is estimated under the two hypotheses, we can derive the GLRT. The log-likelihood ratio (LLR) is:

\small
\begin{align}
	&\Lambda (\bm{z})= \log \frac{f(\bm{z}|\hat{\mathbf{u}}_0,\mathcal{H}_0)}{f(\bm{z}|\hat{\mathbf{u}}_1,\mathcal{H}_1)} =\log f(\bm{z}|\hat{\mathbf{u}}_0,\mathcal{H}_0) -\log f(\bm{z}|\hat{\mathbf{u}}_1,\mathcal{H}_1) \nonumber\\
	&= \frac{1}{2\sigma^2} \sum_{i=1}^{MK}(z[i]-h^{'}\hat{u}_{0}(i))^2-\frac{1}{2\sigma^2} \sum_{i=1}^{MK}(z[i]-h^{'}\hat{u}_{1}(i))^2 \nonumber\\
	&=\Re (\hat{\mathbf{u}}^H_0 \mathbf{z})-\Re (\hat{\mathbf{u}}^H_1 \mathbf{z})  \lessgtr^{\mathcal{H}_1}_{\mathcal{H}_0} 2\sigma^2 \ln \gamma \nonumber\\
	& \Rightarrow T(\bm{z}) \lessgtr^{\mathcal{H}_1}_{\mathcal{H}_0}  =
		\gamma^{'} \label{eqn:GLRT}
\end{align}
\normalsize
Note in the above that $\gamma$ is 1 for the maximum likelihood (ML) criterio. So the sufficient statistic $T(\bm{z})$ is the result of matched filtering the received signal with the two candidate exponentials (with CFO or not) that include also the estimated missing parameters.

\begin{figure}[t]
	\centering	
		\includegraphics[width=0.99\linewidth]{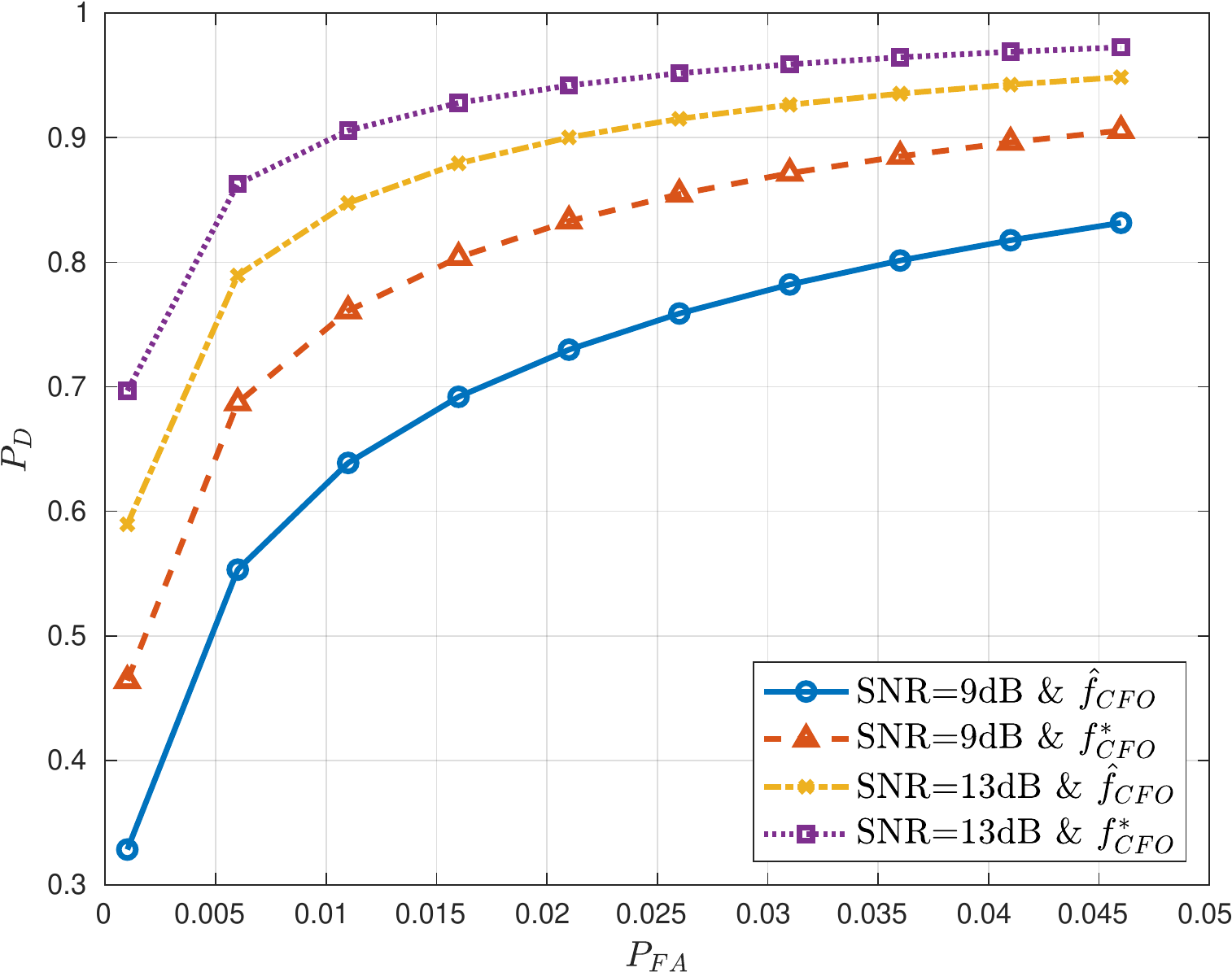}
	\caption{$P_D$ versus $P_{FA}$ for different SNR and different ways of estimating the CFO (genied-aided and the proposed method).}
	\label{fig:PD_PA}
\end{figure}

\section{Simulations \& Discussion}
The objective of our simulations is to evaluate the ability of the detector to detect false targets with high accuracy so that its RADAR functionality is not compromised. We considered an 802.11 OFDM system of 52 carriers in a \SI{20}{\mega\hertz} channel, subcarrier spacing of $\Delta f$=\SI{312.5}{\kilo\hertz}, and 12 sub-carriers were used for pilot signals. We process a number of $M$=10 consecutive OFDM symbols for producing the Range-Doppler response, solving the estimation problem, and finishing with the detection. We set the adversary at an initial distance of $R_0$=\SI{100}{\meter} and the JRC system is moving towards it with a velocity of \SI{10}{\meter\per\second}. Our results were obtained by averaging more than 1000 simulation runs for each threshold $\gamma$ that we tested. We set initially $\gamma$ to a  high negative value, that ensures low false alarm (FA) rate for $\mathcal{H}_1$, while we gradually increased it closer to zero leading to higher FA rate and higher detection probability. The signal-to-noise ratio (SNR) refers to the power of the useful signal in our models versus the power of the noise at the JRC receiver.

In Fig.~\ref{fig:PD_PA} we present results for the detection probability ($P_D$) versus the false alarm rate ($P_{FA}$) for the previous settings. 
We notice that for \SI{9}{\decibel} there is nearly a \SI{10}{\percent} loss in $P_D$ when our system uses the estimator versus a genie-aided system that assumes perfect knowledge of the CFO (denoted as $f^*_\text{CFO}$). With the genie-aided system the detection problem is executed under the ideal match filtering conditions. Performance differences are minimized in higher SNR of \SI{13}{\decibel} since CFO estimation accuracy is improved significantly allowing us to have a better tradeoff between $P_D$ and $P_{FA}$. We expect that our scheme will be more useful in scenarios where a high $P_D$ is necessary while a higher $P_{FA}$ is not so critical, e.g. the ADAS vehicular scenario we described in the Introduction.

One important detail that we can discuss before we conclude is that the adversary can use a technique for tracking the $f_\text{fco}$ (e.g. with a phased-lock loop (PLL)) and try to compensate for it before transmitting the signal. However, this can be easily alleviated if the JRC transmitter adds additional frequency offset that cannot be easily tracked by the PLL at the receiver of the adversary~\cite{jnl_2020_phy}.

\section{ Conclusions}
In this paper we presented a new approach for detecting false targets that are generated from a wireless communication system and intend to confuse  an OFDM-based JRC system. Our method enhances the existing OFDM RADAR algorithm that estimates range and Doppler so that it also estimates CFO by considering the frequency shift that affects each subcarrier. Subsequently, we use the GLRT for defining the detection problem and obtaining effectively a frequency detector. Our results indicate that the probability of detecting false targets is very high with our system. The final result is a scheme that offers the ability to detect falsely generated targets in OFDM-based RADAR systems and it can be easily implemented on top of existing algorithms.

%\section{References}
\bibliographystyle{IEEEtran}
\bibliography{../../../../../tony-bib}

\end{document}